\documentclass[12pt]{iopart}

\usepackage{iopams}
\usepackage{graphicx}
\usepackage{xcolor}
\usepackage{hyperref}

\newlength{\wdth}

\begin{document}
\title{New regime of the Coulomb blockade in quantum dots}

\author{G McArdle$^1$, R Davies$^{1,2}$, I V Lerner$^1$, I V Yurkevich$^2$}

\address{$^1$ School of Physics and Astronomy, University of Birmingham, Birmingham, B15 2TT}
\address{$^2$ School of Computer Science and Digital Technologies, Aston University, Birmingham, B4 7ET}
\ead{i.v.lerner@bham.ac.uk}

\begin{abstract}
{We consider how the absence of thermalisation affects the classical Coulomb blockade regime in quantum dots. By solving the quantum kinetic equation in the experimentally accessible regime when the dot has two relevant occupation states, we calculate the current-voltage characteristics for arbitrary coupling to the leads. If the couplings are strongly asymmetric, we have found that the Coulomb staircase practically reduces to the first step independent of the charging energy when the latter is larger than the Fermi energy, while the  standard thermalised results  are recovered in the opposite case. If the couplings are of the same order, the absence of thermalisation has a new, striking signature --  a robust additional peak in the differential conductance.}
\end{abstract}

\submitto{\JPCM}
\maketitle

\section{Introduction}

Since their discovery, quantum dots have offered insight into a multitude of fundamental transport phenomena in mesoscopic systems \cite{Few_electron_review, QDElectronStructure, Aleiner_Review, Alhassid_Review, Kouwenhoven1997}. The ability to fine-tune their optical and electrical properties means they have also found use in a wide variety of applications \cite{Kouwenhoven1997, QDApplicationReview}. Electronically, the confinement of electrons onto a central island leads to a number of interesting properties, with one of the most well-studied being the Coulomb blockade. Many facets of this regime are now understood and experimentally verified (see \cite{Aleiner_Review, Alhassid_Review, Kouwenhoven1997} for reviews) and it occurs due to electron-electron interactions generating a capacitance for the dot. This results in the presence of a charging energy, given by $E_{\mathrm{c}} = e^2/C$ for a dot of capacitance $C$, which details the energy required to overcome the strong Coulomb interaction on the dot. For large charging energies this leads to the classical Coulomb blockade regime, defined by the separation of energy scales \cite{Kouwenhoven1997}

\begin{equation}\label{scales}
 	\hbar \Gamma \ll \Delta \ll k_B T \ll E_\mathrm{c},
\end{equation}
where $\Gamma$ is the coupling to the leads, and $\Delta$ is the typical energy level spacing between states on the dot that are thermally smeared into a continuum of states by temperature $T$. The rest of this paper will set the Boltzmann and reduced Planck constant to equal one, $\hbar, k_B = 1$.

The defining features of the Coulomb blockade reside in the current-voltage characteristics upon varying both the gate voltage, which controls the preferred number of electrons on the dot, $N_\mathrm{g}$, and the bias voltage across the system. The gate voltage characterises the phenomenology of the equilibrium Coulomb blockade and results in peaks in the conductance that occur at voltages separated by $E_{\mathrm{c}}$ \cite{Kulik, AvLik_Paper, Beenakker}. The peaks occur when the energies of having $N$ and $N{+}1$ electrons on the dot are tuned to degeneracy, resulting in current being able to transfer across the system without an energetic cost. Increasing bias voltage when the couplings to the leads are imbalanced creates an accumulation of electrons in the central island. This is reflected in a distinctive staircase in the current-voltage characteristics known as the Coulomb staircase \cite{Kulik, Averin-Likharev_book_chapter, Ben-Jacob_Wilkins}. If the coupling to the leads are of the same order, then electrons cannot as easily accumulate due to more frequent tunnelling off of the dot. This leads to the staircase becoming less pronounced with a residual signature appearing in the differential conductance \cite{Averin-Likharev_book_chapter, NazarovBlanterBook}. 


Historically, the main approach to the analysis of this regime was through utilising the classical master equation which is justified under the assumption of full thermalisation on the dot. Full thermalisation implies that the thermalisation time, $1/\gamma$, is much larger than the escape time, $1/\Gamma$. The thermalisation rate $\gamma(\varepsilon)$, for large quasiparticle energy $\varepsilon$, is given by \cite{AKGL, Sivan_Imry_1994, Blanter_rates}
\begin{equation}\label{QP_Lifetime}
	\gamma(\varepsilon) \approx \Delta\left(\frac{\varepsilon}{E_\mathrm{Th}}\right)^2,
\end{equation}
where $E_\mathrm{Th} = g\Delta$ is the Thouless energy and $g \gg 1$ is the dimensionless conductance of the dot.


In this paper, we consider the opposite limit when thermalisation is weak, $\gamma \ll \Gamma$, which is experimentally accessible in the classical Coulomb blockade regime (\ref{scales}). Previous work in this non-thermal regime has produced analytical results in linear response \cite{Beenakker} and the numerical results \cite{Averin_Korotkov} for the case of asymmetric coupling to the left and right leads, $\Gamma_L\ne  \Gamma_R$. When the couplings are of the same order, $\Gamma_{\rm L} \sim \Gamma_{\rm R}$, we find a clear signature of the absence of thermalisation in the appearance of a robust extra peak in the differential conductance. 

In addition to the effect of thermalisation, we identity a new regime for weakly populated dots where $T\ll\varepsilon_{\rm F} \ll E_{\rm c}$. This requirement together with (\ref{scales}) describes the regime where the Coulomb staircase (found when $\Gamma_L \gg \Gamma_R$)  becomes virtually unobservable, with the initial step being    $N$ times higher than  the subsequent ones which are practically smeared out. Noticeably, this means that in this regime the first (and only) step is no longer proportional to $E_{\rm c}$.

We have used the Keldysh formalism to reduce the problem to the detailed balance equation, describing the tunnelling processes at a given energy. We present an exact solution in the case where there are only two relevant occupation states of the dot, $N$ and $N{+}1$ (with $N \gg 1$), which is justified under condition (\ref{scales}).

\section{Model}

The standard model for a zero-dimensional dot in the Coulomb blockade regime is described by the Hamiltonian \cite{Aleiner_Review, Alhassid_Review,Kouwenhoven1997},
\begin{equation}\label{H_dot}
    H_\mathrm{d} = \sum_n \varepsilon_n d_n^\dagger d_n + {\textstyle{\frac{1}{2}}} {E_\mathrm{c}} \left(\hat{N} - N_\mathrm{g} \right)^2,
\end{equation}
where $d_n^\dagger \left(d_n\right)$ are the creation (annihilation) operators of the quantum dot for the level $n$ with energy $\varepsilon_n$ and $\hat{N} = \sum_n d_n^\dagger d_n$ is the number operator for the dot. It is useful to introduce a variable,$\Omega_N$, that characterises the difference in interaction energies between the $N$ and $N{+}1$ states which is defined to be
\begin{equation}
    \Omega_N = E_{\mathrm{c}}\Big(N+{\textstyle{\frac{1}{2}}}-N_{\mathrm{g}}\Big).
\end{equation}

In order to study the $I$-$V$ characteristics, the central island is coupled to left ($\mathrm{L}$) and right ($\mathrm{R}$) leads. The Hamiltonian of the individual leads, $H_\mathrm{l}$ and the tunnelling between them and the dot, $H_\mathrm{t}$, are given by,
\begin{equation}\label{H_l}
    H_\mathrm{l}  = \sum_{k, \alpha} \left(\varepsilon _{k}-\mu _\alpha\right ) c_{k, \alpha}^\dagger c_{k,\alpha},
\end{equation}
\begin{equation}\label{H_t}
	H_\mathrm{t} = \sum_{\alpha, k, n} \left( t_{\alpha } c_{k,\alpha}^\dagger d_n + \mathrm{h.c.} \right).
\end{equation}
Therefore, these terms constitute the Hamiltonian of the entire system
\begin{equation}\label{H}
    H = H_\mathrm{d} + H_\mathrm{l}  + H_\mathrm{t}.
\end{equation}
In the above $c_{k,\alpha}^\dagger \left(c_{k, \alpha}\right)$ are the creation (annihilation) operators for an electron in lead $\alpha=\{\mathrm{L}, \mathrm{R}\}$. These electrons have an energy $\left(\varepsilon _{k}-\mu _\alpha\right )$, with the chemical potentials of the leads given in this work by $\mu_\mathrm{L} = \mu +eV$ and $\mu_\mathrm{R} = \mu$ {where $V$ is the applied source-drain voltage across the dot. Different ways of allocating the voltage to the two leads can be considered and results suitably generalised.} The tunnelling amplitude $t_{\alpha }$ is assumed to be independent of $k$ and $n$ and defines the broadening of the energy levels of dot caused by the presence of the leads, $\Gamma$. Taking the density of states in the leads $\nu_\alpha$ to be a constant, the coupling of the dot to lead $\alpha$ is given by $\Gamma_\alpha = 2 \pi \nu_\alpha |t_\alpha|^2$, with the total coupling $\Gamma = \Gamma_{\mathrm{L}} + \Gamma_{\mathrm{R}}$. The asymmetry ratio of $\Gamma_\mathrm{L}/\Gamma_\mathrm{R}$ will be of particular importance in this work.

\section{Method}

Previous analytical attempts to understand linear response in the Coulomb blockade regime, albeit for $\Delta \gg T$, given the absence of inelastic processes have resulted in the derivation of a detailed balance equation where at a given energy the tunnelling rates both on and off the dot are equal \cite{Beenakker}. For the non-equilibrium regime, we utilise the Keldysh formalism (see \cite{Rammer_Smith} for a review) and the quantum kinetic equation (QKE) in a way similar to that detailed in \cite{Meir_Wingreen_Jauho}. A similar approach has previously been used to calculate the tunnelling density of states near to equilibrium \cite{Kamenev_Gefen,TDoS}.

The QKE can be written in terms of the probability, $p_N$, that the dot has $N$ electrons on it and the distribution function of the $N$ electron dot, $F_N(\varepsilon_n)$,
\begin{eqnarray}\label{QKE}
    p_N \left(1-F_N(\varepsilon_n)\right)\widetilde {f}(\varepsilon_n + &\Omega_N) &= p_{N+1}F_{N+1}(\varepsilon_n) \left(1-\widetilde {f}(\varepsilon_n + \Omega_N) \right),
\end{eqnarray}
where
\begin{eqnarray}\label{ftilde}\widetilde {f}(\varepsilon) &= \frac{\Gamma_{\mathrm{L}}}{\Gamma}f(\varepsilon-\mu_\mathrm{L}) + \frac{\Gamma_{\mathrm{R}}}{\Gamma}f(\varepsilon-\mu_\mathrm{R}).
\end{eqnarray}
for Fermi function $f(\varepsilon)$. The normalisation conditions are $\sum_{N} p_N=1$ and $\sum_{n} F_N(\varepsilon _n)=  (1/\Delta)\int_0^{\infty} F(\varepsilon)\mathrm{d}\varepsilon=N$. {The details of the derivation, as well as its exact solution in a two-state limit, is presented in \ref{Sec:Derivations}. The QKE is analogous to the detailed balance equations derived in \cite{Beenakker} for $\Delta \gg T$. 
Due to the nature of the Coulomb blockade, the dot will commonly be in a two-state limit where there are only two relevant states, $N$ and $N{+}1$, with all others being exponentially suppressed. With this simplification, the solution to the QKE when $N \gg 1$ can be summarised as $F_N(\varepsilon_n) \approx F_{N+1}(\varepsilon_n) \approx F(\varepsilon_n)$, where}

\begin{equation} \label{QKE F Soln}
	F(\varepsilon_n)  = \frac{\widetilde{f}(\varepsilon_n + \Omega_N)}{[1-\widetilde{f}(\varepsilon_n + \Omega_N)]\frac{p_{N+1}}{p_N}+\widetilde{f}(\varepsilon_n + \Omega_N)}.
\end{equation}
The ratio of probabilities is determined by the normalisation of the distribution function, which fixes the number of particles on the dot $N = \sum_n F(\varepsilon_n)$, and the individual probabilities can be subsequently found from $p_N + p_{N+1}=1$.

Upon calculation of the probabilities and distribution function, it is necessary to see how they manifest in the experimentally observable $I$-$V$ characteristics. Through calculation of the Green's functions of the dot (see \ref{Sec:Derivations}), the standard expression for the tunnelling current through the lead $\alpha$ \cite{Meir_Wingreen_Jauho, Haug_Jauho} is written as, 
\begin{equation}\label{I_lead}
	\fl I_\alpha = e\Gamma_\alpha \sum_N p_N \sum_n \Big(F_N(\varepsilon_n)\left[1-f (\varepsilon_n {-}\mu _\alpha {+} \Omega_{N{-}1})\right]- \left[1-F_N(\varepsilon_n)\right] f (\varepsilon_n {-}\mu _\alpha {+} \Omega_N) \Big).
\end{equation}
Current conservation, $ I=I_{\mathrm{R}}=-I_{\mathrm{L}} $, can be used to finally express the current, where $f_\alpha(\varepsilon_n)=f(\varepsilon_n-\mu_\alpha)$,
\begin{eqnarray}\label{Current}
	\fl I = e\frac{\Gamma_{\mathrm{L}} \Gamma_{\mathrm{R}}}{\Gamma} \sum_N p_N \sum_n \Big( F_N(\varepsilon_n)&\left[f_\mathrm{L} (\varepsilon_n +\Omega_{N-1} ) - f_\mathrm{R} (\varepsilon_n+\Omega_{N-1})\right] \nonumber\\&+ (1-F_N(\varepsilon_n)) \left[f_\mathrm{L}(\varepsilon_n +\Omega_N) - f_\mathrm{R} (\varepsilon_n +\Omega_N) \right] \Big).
\end{eqnarray}

\section{Results and Discussion}

Using the formalism outlined above, the $I$-$V$ characteristics can be calculated in the absence of thermalisation. The completely thermalised results can be obtained by replacing the distribution function with its equilibrium value, $f(\varepsilon_n-\varepsilon_\mathrm{F})$, and then integrating the QKE (\ref{QKE}) over all energies to obtain the probabilities. {The current is then obtained using (\ref{Current}). We explore both asymmetric and symmetric coupling to the leads, identifying a new regime in each situation. Prior to focusing on the current, the probability of having $N$ particles on the dot and its associated distribution function must first be calculated. This will illuminate the origin of the new results.}


\subsection{Probabilities}

Our solution to the QKE is limited to the situation when only two probabilities of occupation are relevant. In the low-temperature equilibrium dynamics of the problem, this condition is always satisfied due to the charging energy being the largest scale in the system. In non-equilibrium however, when the source-drain voltage becomes larger than the charging energy, only a restricted parameter space can be investigated. The probabilities are found from the normalisation of the distribution function, (\ref{QKE F Soln}),
\begin{equation}\label{F Normalisation}
    \varepsilon_\mathrm{F} \equiv N\Delta = \int_0^\infty \frac{\widetilde{f}(\varepsilon_n + \Omega_N)}{[1-\widetilde{f}(\varepsilon_n + \Omega_N)]\frac{p_{N+1}}{p_N}+\widetilde{f}(\varepsilon_n + \Omega_N)},
\end{equation}
in combination with the normalisation of the probabilities, $p_N+p_{N+1} = 1$. In the case where the coupling to the leads is asymmetric, this integral can be performed exactly and gives for $\Gamma_\mathrm{L} \gg \Gamma_\mathrm{R}$
\begin{equation}\label{LeftProbs}
    \frac{p_{N+1}}{p_N} = \mathrm{e}^{-\beta(\varepsilon_\mathrm{F} -\mu_\mathrm{L} +\Omega_N)}.
\end{equation}
In the opposite limit of $\Gamma_\mathrm{L} \ll \Gamma_\mathrm{R}$, we obtain the same result except with the replacement $\mu_\mathrm{L} \rightarrow \mu_\mathrm{R}$. In both instances of strong asymmetry, the results are identical to those in the fully thermalised case \cite{Kulik, Averin-Likharev_book_chapter,Ben-Jacob_Wilkins} and mean that for any bias voltage, $V$, there only at most two relevant states and therefore the current can be calculated for any $V$.

\begin{figure}[h!]
    \centering
    \includegraphics{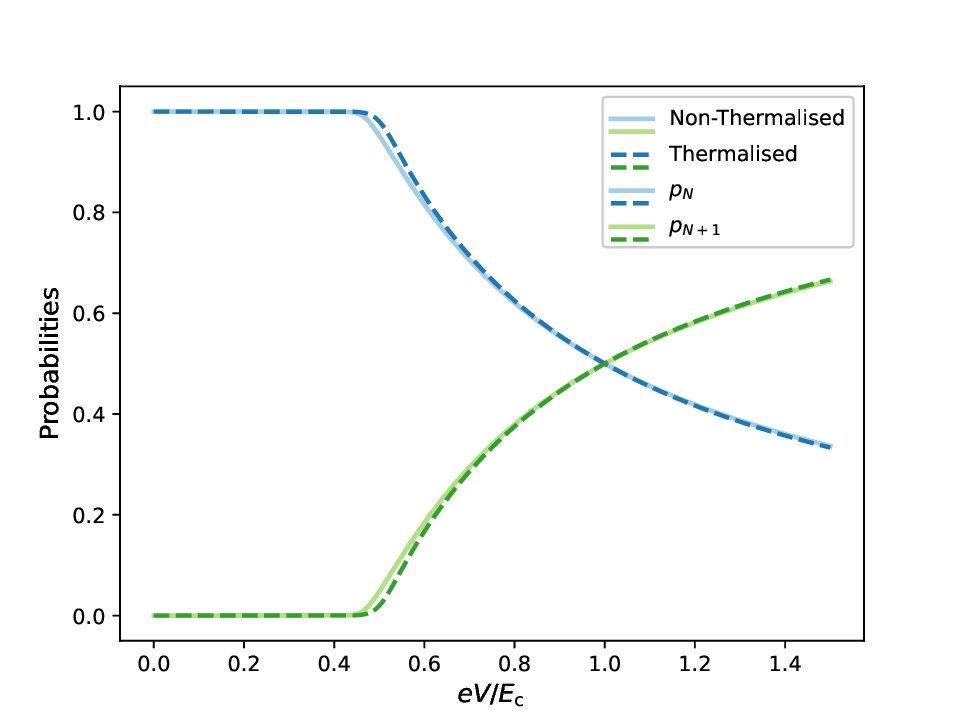}
    \caption{Demonstrating how the probabilities of having $N$ and $N{+}1$ electrons occupying the dot change upon application of source-drain voltage. This is calculated from both the master equation and quantum kinetic equation corresponding to the thermalised (dashed lines) and non-thermalised (solid lines) cases respectively for the case of equal coupling to the leads. These two regimes produce virtually identical probabilities with the small difference around $eV \sim \Omega_N = E_{\mathrm{c}}/2$ becoming smaller as $E_{\mathrm{c}}/T$ increases, with $E_{\mathrm{c}}/T=100$ here.}
    \label{fig:ThermProb}
\end{figure}

In order to consider the case of approximately symmetric couplings, the low-temperature expansion of $\widetilde{f}(\varepsilon_n)$ must be considered,
\begin{equation}
\fl \widetilde {f}(\varepsilon_n +\Omega_N) \approx\  \left\{\begin{array}{ll}
1 -(\Gamma _{\mathrm{R}}/\Gamma) \mathrm{e}^{\beta(\varepsilon_n-(\mu-\Omega_N))}, &\varepsilon_n < \mu-\Omega_N \\
		\Gamma _{\mathrm{L}}/\Gamma,  &\mu-\Omega_N < \varepsilon_n < \mu-\Omega_N+eV \\
		(\Gamma _{\mathrm{L}}/\Gamma) \mathrm{e}^{-\beta[\varepsilon_n-(\mu -\Omega_N+eV)]}, &\mu -\Omega_N+eV<\varepsilon_n. 
\end{array} \color{white}\right\}\color{black}
\end{equation}
Performing the integral in (\ref{F Normalisation}) gives the following equation to be solved numerically for the ratio of probabilities,
\begin{equation}\label{AN Eqn}
	\beta \varepsilon_\mathrm{F} = \frac{\beta e V}{\frac{p_{N+1}}{p_N}\frac{\Gamma_\mathrm{R}}{\Gamma_\mathrm{L}}+1} +\ln \left( \frac{\Gamma}{\Gamma _{\mathrm{R}}}\frac{p_N}{p_{N+1}}\mathrm{e}^{\beta(\mu - \Omega_N)}+1\right)  + \ln \left(\frac{\frac{\Gamma _{\mathrm{L}}}\Gamma  +   \frac{p_{N+1}}{p_N} } {\frac\Gamma{\Gamma _{\mathrm{R}}}  +   \frac{p_{N+1}}{p_N} }   \right).
\end{equation}
The solutions for this are shown in Figure \ref{fig:ThermProb}. We highlight that the probabilities in the non-thermalised regime are extremely similar to the thermalised ones. This insensitivity to thermalisation is caused by the fact that the probabilities are determined almost solely by the energetics of the problem.  For voltages less than $\Omega_N$ (assuming that $\mu = \varepsilon_\mathrm{F}$), the dot is firmly in the blockade region with $N$ electrons on the dot. As the voltage is increased above this threshold, it becomes energetically possible for an electron to enter the dot and both $N$ and $N{+}1$ particle states become relevant. This persists until $eV \approx \Omega_{N+1}$ when another electron could enter the dot, violating the two-state condition that predicates our solution. In the region of validity, (\ref{AN Eqn}) can be used to estimate that $p_N = p_{N+1}$ occurs at $eV \approx (1 + \frac{\Gamma_\mathrm{R}}{\Gamma_\mathrm{L}})(\varepsilon_\mathrm{F}-\mu+\Omega_N)$.

\subsection{Distribution Function}

Following the calculation of the probabilities, the distribution function is easily obtained by substituting the result for the ratio, $p_{N+1}/p_N$ into (\ref{QKE F Soln}). In the case of asymmetric coupling to the leads (regardless of which lead is more strongly coupled), the distribution function becomes a Fermi function with chemical potential $\varepsilon_\mathrm{F}$. Therefore, in the non-thermalised case a strong asymmetry of the coupling reproduces the behaviour of the thermalised dot as the coupling effectively causes an equilibration across all energies on the dot.

\begin{figure}
    \centering
    \includegraphics{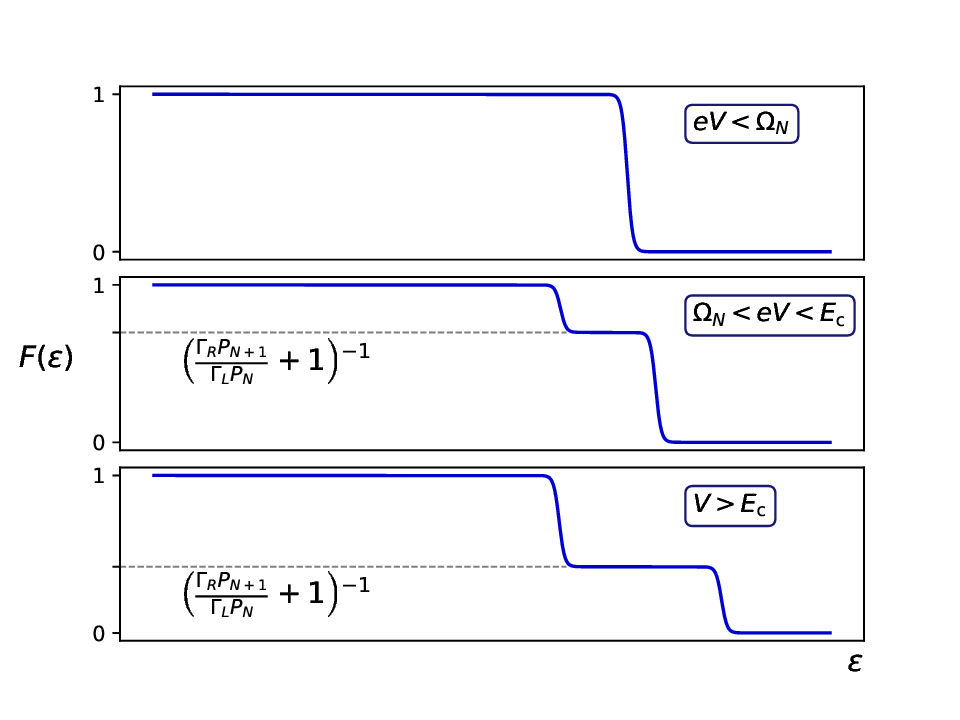}
    \caption{The distribution function of the electrons on the dot shown for three different voltages. For voltages less than $\Omega_N$, an incoming electron cannot overcome the charging energy and the distribution remains as a Fermi function. At larger voltages the double step appears and widens at higher bias. The height of the additional step decreases as voltage increases due to the dependence of $p_{N+1}/p_N$ on the source-drain voltage.  }
    \label{fig:NonEqDist}
\end{figure}

On the other hand, when the couplings to both leads are approximately equal, the non-equilibrium distribution of the dot is profoundly different from the equilibrium result, as there is a lack of equilibration between electrons from the left and right leads. A distinctive double-step feature emerges as higher energy states can be occupied by electrons from the left lead,
\begin{equation}\label{FN regions}
	F(\varepsilon_n) \approx\ \left\{ \begin{array}{ll}
		1, &  \varepsilon_n {<} \mu_{\mathrm{R}}{-}\Omega_N\\
		\left(1{+}\frac{\Gamma_\mathrm{R}}{\Gamma_\mathrm{L}}\frac{p_{N+1}}{p_N} \right)^{\!{-}1}\!\!\!,  &\mu_{\mathrm{R}}{-}\Omega_N   {<}\varepsilon_n {<} \mu_{\mathrm{L}}{-}\Omega_N \\
		0, & \mu_{\mathrm{L}} {-}\Omega_N {<}\varepsilon_n  
\end{array} \color{white}\right\} \color{black}
\end{equation}
which is displayed for different voltages in Figure \ref{fig:NonEqDist}. A similar change also occurs for dots involving non-interacting electrons \cite{AltlandSimons, AltlandEgger, SmirnovKondoQD} or a one-dimensional wire \cite{Birge}, where the distribution function is a linear combination of those of the leads. However, in this instance, the double-step form is not as simple and is significantly modified by the interaction.

This distribution function is the source of differences in behaviour between thermalising and non-thermalising quantum dots. Its form is dependent on the scales of parameters in the system. For small voltages, $eV< \Omega_N$, which are insufficient to compensate the charging energy, there are $N$ electrons on the dot and the distribution is a Fermi function. As the voltage is increased, such that two states become relevant, the double-step form becomes relevant. The height of the middle step is given by $\left(1{+}\frac{\Gamma_\mathrm{R}}{\Gamma_\mathrm{L}}\frac{p_{N+1}}{p_N} \right)^{\!{-}1}$, meaning that asymmetry in the coupling will dampen the effect of the non-equilibrium behaviour. Indeed, if one of the couplings is taken to zero then the asymmetric results previously mentioned are important and the thermalised results are recovered.

\subsection{Current-Voltage Characteristics}

Having obtained the distribution function of the dot and probabilities of occupation, the current-voltage characteristics can now be obtained using (\ref{Current}). Beginning with the asymmetric coupling, we first analyse the case of a large Fermi energy on the dot ($\varepsilon_\mathrm{F} \gg E_\mathrm{c}$). As previously stated, in the asymmetric case, there are never more than two relevant states so the current can be obtained for all voltages. Using that $F_N(\varepsilon_n) = f(\varepsilon_n{-}\varepsilon_\mathrm{F})$ and the result in (\ref{LeftProbs}) with the appropriate normalisation, $\sum_N p_N = 1$, the current when $\Gamma_\mathrm{L} \gg \Gamma_\mathrm{R}$ is found from (\ref{Current}) to be,
\begin{eqnarray}
	I = 0, & \qquad 0 &\leq eV \lesssim \Omega_{N_0} \qquad \quad   (p_{N_0} = 1), \nonumber
\\[4pt]
	I = e\Gamma_{\mathrm{R}} \frac{\Omega_{N_0}  }{\Delta} , & \quad \Omega_{{N_0}}   &\lesssim eV\lesssim \Omega_{{N_0}+1}  \qquad  (p_{{N_0}+1} = 1),\label{Large_Results}\\[4pt]
	 I = e\Gamma_{\mathrm{R}} \frac{\Omega_{{N_0}+1}  }{\Delta} ,\qquad &\Omega_{{N_0}+1}   &\lesssim eV\lesssim \Omega_{{N_0}+2}     \qquad(p_{{N_0}+2} = 1), \nonumber
 \end{eqnarray}
and so on. 
Here, $N_0$ is the number of electrons in equilibrium. These results are identical to those in the case of complete thermalisation \cite{Kulik, Averin-Likharev_book_chapter, Ben-Jacob_Wilkins}, reflecting the fact that the strong asymmetry in the coupling leads to the distribution function on the dot taking its equilibrium form.  The staircase present in the current here exists as electrons can accumulate on the dot. In the opposite limit of $\Gamma_\mathrm{R} \gg \Gamma_\mathrm{L}$, this will no longer be true as the bias is applied to the left lead only and therefore the current is simply Ohmic for $eV>\Omega_{N_0}$.

Although for a large Fermi energy, the absence of thermalisation has no impact in the strongly asymmetric case, we identify a new parametric regime when the Fermi energy of the dot is much less than the charging energy. In obtaining the above results it is important to note that the sum over energy levels is converted to an integral through $\sum_{n} \rightarrow  (1/\Delta)\int_0^{\infty} \mathrm{d}\varepsilon$, where we note the lower integration limit accounts for the bottom of the dot. In the case when $\varepsilon_\mathrm{F} \ll E_\mathrm{c}$ this becomes vital as now $\varepsilon_\mathrm{F} < \Omega_{N_0}$ is possible ($\Omega_{N_0} \approx E_\mathrm{c}/2$ in the middle of the valley). Taking care with the integration (as is also described in \cite{AsymCBPaper}), the current is given by
\begin{eqnarray}
	I = 0, & \qquad 0 &\leq eV \lesssim \Omega_{N_0} \qquad \quad   (p_{N_0} = 1), \nonumber
\\
\label{Small_Results}I = e\Gamma_{\mathrm{R}} ({N_0}+1),   & \quad \Omega_{{N_0}}   &\lesssim eV\lesssim \Omega_{{N_0}+1}  \qquad  (p_{{N_0}+1} = 1),\\
	I = e\Gamma_{\mathrm{R}} ({N_0}+2)   ,\qquad &\Omega_{{N_0}+1}   &\lesssim eV\lesssim \Omega_{{N_0}+2}     \qquad(p_{{N_0}+2} = 1), \nonumber
\end{eqnarray}
and so on. We wish to highlight that the step heights are no longer proportional to $E_\mathrm{c}/\Delta$, unlike the large Fermi energy results, as the lowest energy levels of the dot now dominate the transport. Although at an initial glance it appears that a staircase persists, we note that the first step is proportional to $N_0 \gg 1$. Therefore this step is much larger than the subsequent steps and therefore the staircase practically vanishes.

\begin{figure}
    \centering   \includegraphics{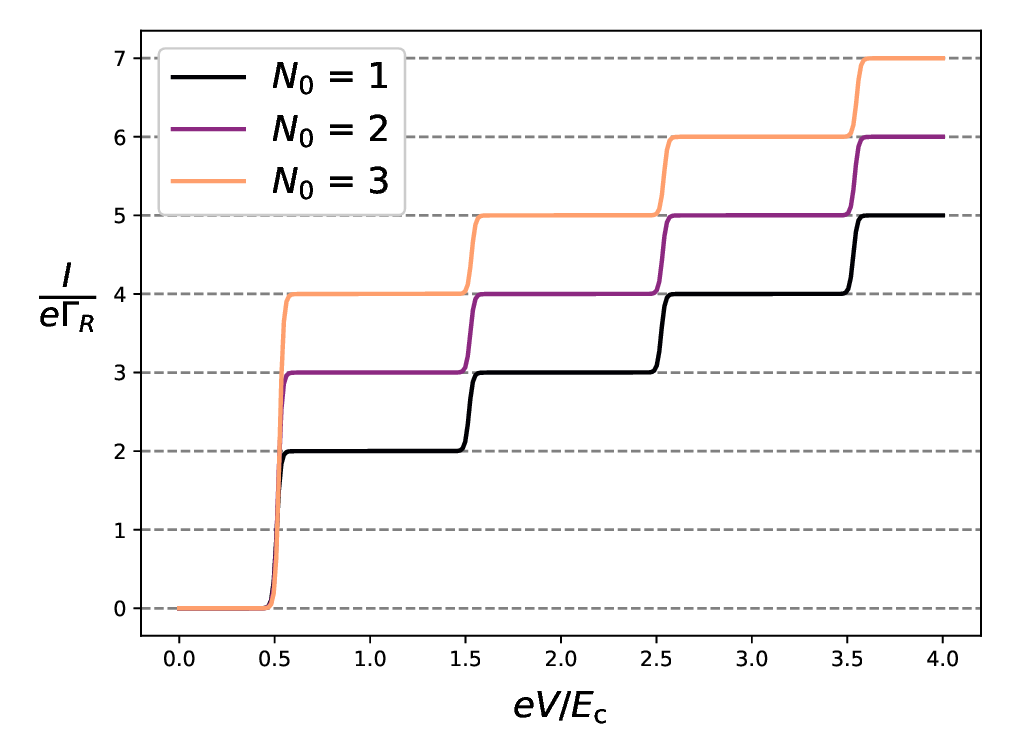}
    \caption{The initial number of particles on the dot, $N_0$, determines the size of the first step in the current in the asymmetric regime for $\varepsilon_\mathrm{F} \ll E_\mathrm{c}$ with subsequent steps being independent of $N_0$. Here the current is found exactly for a $7$ state dot with thermally smeared energy levels.   }
    \label{fig:AsymmSmallNg}
\end{figure}

This regime also corresponds to a stepping stone between the classical and quantum blockade regimes. The quantum regime is defined by Eq (\ref{scales}) but with $\Delta \gg T$ such that the individual energy levels can be distinguished. To verify this we solve the quantum master equation \cite{QmeqPackage} numerically for a dot with 7 states, with the results shown in Figure \ref{fig:AsymmSmallNg}. We find the same $I$-$V$ curves, suggesting that our results in (\ref{Small_Results}) persist down to small values of $N\sim 10$, therefore bridging the gap between the classical Coulomb blockade and the quantum regime where smaller numbers of electrons are on the dot and the low-lying energy levels dominate the transport (see, for example, \cite{Few_electron_review}). 

\begin{figure}
    \centering
    \includegraphics[scale=0.9]{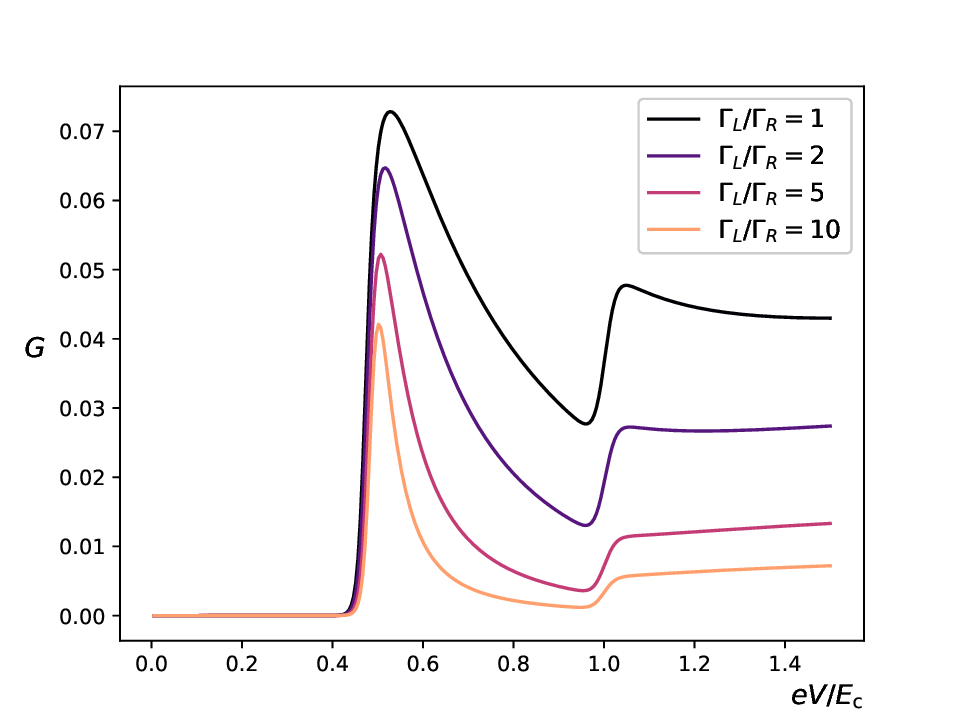}
    \caption{The differential conductance is plotted as a function of bias voltage in the case of $\varepsilon_\mathrm{F} \gg E_\mathrm{c}$ for varying asymmetry of the coupling to the leads. The secondary jump in conductance at $eV = E_{\mathrm{c}}$ is the distinctive feature of non-thermalising quantum dots. This becomes smaller, in comparison to the first peak, with increasing asymmetry.}
    \label{fig:ChangingGammaR}
\end{figure}

The absence of thermalisation has a more pronounced impact for symmetric coupling with $\varepsilon_\mathrm{F} \gg E_\mathrm{c}$ despite there being a smaller accessible voltage range, $eV \lesssim \Omega_{N_0+1}$, due to the restriction to two relevant states. This is due to the changing of the distribution function from a Fermi function to the double-step form in (\ref{FN regions}). The key change this leads to is an additional jump in the differential conductance at $eV = E_{\mathrm{c}}$ as shown in Figure \ref{fig:ChangingGammaR}. {The position of this jump is robust due to $\Omega_{N_0+1}-\Omega_{N_0} = E_\mathrm{c}$. }Importantly, this jump is experimentally observable with it's height given by (for $\Omega_{N_0} = E_\mathrm{c}/2$),
\begin{equation}\label{G jump}
	\delta G = \frac{e^2}{2\Delta}\frac{\Gamma_\mathrm{L}\Gamma_\mathrm{R}}{\Gamma}.
\end{equation}
In order to calculate this we note that around the jump, the ratio of probabilities is (also see \cite{McADLYPRL})
\begin{equation}\label{Jump AN}
	\frac{p_{N+1}}{p_N}\approx \frac{\Gamma_\mathrm{L}}{\Gamma_\mathrm{R}}\left(\frac{eV-\Omega_{N_0}}{\Omega_{N_0}} \right),
\end{equation}
and then the current (and therefore the differential conductance) can be obtained on either side of the jump, so that (\ref{G jump}) can be found.

\section{Conclusion}\label{Sec:Discussion}

To summarise, through the calculation of the $I$-$V$ characteristics, we have been able to identify new regimes that are relevant to the classical Coulomb blockade, (\ref{scales}), in the absence of thermalisation. For strongly asymmetric coupling to the leads, both the probabilities of occupation and the distribution function maintain their thermalised form, as a form of equilibrium is established with the more strongly coupled lead. In the limit of a large Fermi energy on the dot, this reproduces the standard Coulomb staircase \cite{Kulik, Averin-Likharev_book_chapter, Ben-Jacob_Wilkins}. However, when the Fermi energy is smaller than the charging energy of the dot the staircase is no longer observable as the first step washes out the subsequent smaller steps for a large number of electrons on the dot. 

In the case of symmetric coupling to the leads the distribution acquires a double-step form that is heavily influenced by the interaction, whilst the probabilities remain close to their thermalised counterparts. This change in distribution function reflects the absence of equilibration of energies between electrons coming from the different leads and those already on the dot. It leads to an additional peak in the differential conductance at voltages equal to the charging energy and should, in principle allow for the detection of whether a quantum dot is in the non-thermalising regime.

\section*{Acknowledgements}
We gratefully acknowledge  support from EPSRC  under the grant EP/R029075/1 (IVL) and   from the Leverhulme Trust under the grant  RPG-2019-317 (IVY).

\section*{References}

\bibliographystyle{prsty}
\bibliography{bibliography}

\newpage
\appendix

\section{Exact Solution in the Two-state Regime}\label{Sec:Derivations}

The method used to obtain the discussed results can be generalised to similar tunnelling problems and here we derive the QKE (\ref{QKE}) in the non-equilibrium regime and demonstrate how to solve it in the case where there are only two relevant occupations of the dot, $N$ and $N{+}1$, with all other states being exponentially suppressed due to the presence of the charging energy.

\subsection{Deriving the QKE}

For an isolated dot the Hamiltonian for the system is simply given by $H_\mathrm{d}$. The Green's function for a single level can be expressed in terms of the time-independent operator, $d_n(t) = \mathrm{e}^{iH_\mathrm{d}t}d_n\mathrm{e}^{-iH_\mathrm{d}t}$, 
\begin{equation}\label{GF_Defn}
	g_n^>(t) = -i \Tr\left(\hat{\rho}_0 d_n(t)d_n^\dagger\right), \hspace{5pt} g_n^<(t) = i \Tr\left(\hat{\rho}_0 d_n^\dagger d_n(t)\right),
\end{equation}
where $\hat{\rho}_0$ is the density matrix for the isolated system. The full Green's function for an isolated dot is then given by a sum over all levels, $n$. Additionally, since the number of electrons on the dot is conserved in the absence of tunnelling, we write (\ref{GF_Defn}) as a sum over subspaces where the number of electrons, $N$ is fixed, 
\begin{eqnarray}\label{g>_isolated}
    \fl g^>_n(\varepsilon) = -2\pi i \sum_N \delta \left(\varepsilon - \varepsilon_n - \Omega_N \right) g^>_N(\varepsilon_n), \hspace{10pt} &g^>_N(\varepsilon_n) = \Tr_N \left(\hat{\rho}_0 d_n d_n^\dagger \right), \\
	\label{g<_isolated}
    \fl g^<_n(\varepsilon) = -2\pi i \sum_N \delta \left(\varepsilon - \varepsilon_n - \Omega_{N-1} \right) g^<_N(\varepsilon_n), \hspace{10pt} &g^<_N(\varepsilon_n) = -\Tr_N \left(\hat{\rho}_0 d_n^\dagger d_n \right).
\end{eqnarray}
We express $g^>_N(\varepsilon_n)$, $g^<_N(\varepsilon_n)$ using more natural parameters to describe the system, that is the probability of having $N$ electrons on the dot, $p_N$, and the distribution function of the dot given that it has $N$ electrons, $F_N(\varepsilon_n)$ via the ansatz
\begin{equation}\label{ansatz}
    g^>_N(\varepsilon_n) = p_N\left(1-F_N(\varepsilon_n)\right)\quad \mathrm{ and }\quad g^<_N(\varepsilon_n) = -p_N F_N(\varepsilon_n).
\end{equation}
This leads to the normalisation of (\ref{g>_isolated}, \ref{g<_isolated}) becoming $\sum_N\left(g^>_N(\varepsilon_n) - g^<_N(\varepsilon_n) \right) =\sum_N p_N = 1$.

Given our separation of scales in (\ref{scales}), we incorporate the effect of the tunnelling to and from non-interacting leads in the weak-coupling limit, $\Gamma \rightarrow 0$. The associated quantum kinetic equation (QKE) is therefore,
\begin{equation}\label{QKE_original}
	g^{>, <}_n \left(\varepsilon\right) = g^{\mathrm{R}}_n \left(\varepsilon\right) \Sigma^{>, <} \left(\varepsilon\right) g^{\mathrm{A}}_n \left(\varepsilon\right),
\end{equation}
with the self-energies given by their standard expressions \cite{Meir_Wingreen_Jauho, Haug_Jauho}
\begin{eqnarray}\label{Self-energy>}
    \Sigma^>(\varepsilon) &=& = - i \left[\Gamma - \left(\Gamma_{\mathrm{L}} f_{\mathrm{L}}(\varepsilon) + \Gamma_{\mathrm{R}} f_{\mathrm{R}}(\varepsilon) \right) \right],
\\\label{Self-energy<}
    \Sigma^<(\varepsilon) &=&  = i\left(\Gamma_{\mathrm{L}} f_{\mathrm{L}}(\varepsilon) + \Gamma_{\mathrm{R}} f_{\mathrm{R}}(\varepsilon)\right).
\end{eqnarray}
The self-energies here are assumed to be independent of the level $n$ and are given in terms of the Fermi functions of the leads, $f_\alpha(\varepsilon) = f(\varepsilon-\mu_\alpha)$. After inserting the self-energies into (\ref{QKE_original}) and rewriting the QKE as $g_n^>(\varepsilon)\Sigma^<(\varepsilon) = g_n^<(\varepsilon) \Sigma^>(\varepsilon)$, we find the result presented in (\ref{QKE}).

\subsection{Exact Solution to the Quantum Kinetic Equation}

In order to see how the solution (\ref{QKE F Soln}) arises, we present an exact solution to the QKE (\ref{QKE}) when there are only two relevant states,
\begin{eqnarray}\label{ExactResult}\nonumber
	p_N &= \frac{Z_N}{Z_N+Z_{N+1}}, \qquad \qquad p_{N+1} &= \frac{Z_{N+1}}{Z_N+Z_{N+1}}, \\[-9pt]\label{Full_Soln}\\[-9pt]
	F_N(\varepsilon_n) &= \frac{Z_N(\varepsilon_n)}{Z_N}, \qquad \qquad F_{N+1}(\varepsilon_n) &= \frac{Z_{N+1}(\varepsilon_n)}{Z_{N+1}}, \nonumber
\end{eqnarray}
where
\begin{eqnarray}
 	Z_N &= \sum_{\{n_j=0,1\}} \prod_{j=1}^\infty \left[\frac{\widetilde {f}(\varepsilon_j + \Omega_N)}{1 - \widetilde {f}(\varepsilon_j + \Omega_N)}\right]^{n_j} \delta_{(\sum_j n_j), N},
\nonumber \\[-9pt]\label{Z_defn} \\[-9pt]\nonumber
 	Z_{N+1} &= \sum_{\{n_j=0,1\}}  \prod_{j=1}^\infty \left[\frac{\widetilde {f}(\varepsilon_j + \Omega_N)}{1 - \widetilde {f}(\varepsilon_j + \Omega_N)}\right]^{n_j} \delta_{(\sum_j n_j), N+1},
\end{eqnarray}
while $Z_{N}(\varepsilon_n)$ in (\ref{Full_Soln}) is defined by restricting the sums in (\ref{Z_defn})  to configurations where $\varepsilon_n$ is occupied. By expressing the Krönecker delta as an integral over $\theta$, we express $Z_N$ as
\begin{equation}\label{Z_saddle}
	Z_N = \int \frac{\mathrm{d}\theta}{2\pi} \mathrm{e}^{Nf(\theta)}, \qquad f(\theta) = \frac{1}{N}\sum_j \ln\left( 1+ \frac{\widetilde {f}(\varepsilon_j + \Omega_N)}{1 - \widetilde {f}(\varepsilon_j + \Omega_N)}\mathrm{e}^{i\theta}\right) -i\theta,
\end{equation}
and as we consider $N \gg 1$, then this is evaluated using the saddle point method to give $Z_N = g(\theta_0)\mathrm{e}^{-iN\theta_0}$. The saddle point equation for the optimal $\theta = \theta_0$ is simply the normalisation of the distribution function given in (\ref{QKE F Soln}). The $N$ dependence of $\theta_0$ is only via $N\Delta$ as both $Z_N$ and $Z_{N+1}$ depend on $\Omega_N$ (rather than $\Omega_{N+1}$) and therefore $\theta_0$ is the same for both (as $N \gg 1$) and so the solutions presented in (\ref{ExactResult}) are
\begin{equation}\label{p and F}
	\frac{p_{N+1}}{p_N} = \mathrm{e}^{-i\theta_0}, \hspace{10pt} F_N(\varepsilon_n)   \approx F_{N+1}(\varepsilon_n) \approx \left({\frac{1 - \widetilde {f}(\varepsilon_n + \Omega_N)}{ \widetilde {f}(\varepsilon_n + \Omega_N)}\mathrm{e}^{-i\theta_0}+1}\right)^{-1},
\end{equation} 
which is equivalent to (\ref{QKE F Soln}).

\end{document}